\documentclass[a4paper,11pt]{article}
\begin{document}

\title{ On the resolution of the big bang singularity in isotropic Loop 
Quantum Cosmology}

\author{Madhavan Varadarajan\\{\sl Raman Research Institute, Bangalore 560 080}\\
madhavan@rri.res.in}




\maketitle

\begin{abstract}
In contrast to previous work in the field, we construct the Loop Quantum
Cosmology (LQC) of the flat isotropic model with a massless scalar field 
in the absence of higher order curvature corrections to
the gravitational part of the Hamiltonian constraint.
The matter part of the constraint contains the inverse triad operator which
 can be quantized with or without the use of 
a Thiemann- like procedure. With the 
latter choice, we show that the LQC quantization is identical
to that of the standard  Wheeler DeWitt theory (WDW) 
wherein there is no singularity resolution. 
We argue that the former choice leads to singularity resolution in the 
sense of a well defined, regular (backward) evolution through and beyond
the epoch where the size of the universe vanishes. 

Our work along with that of the seminal work of Ashtekar, Pawlowski and Singh (APS)
clarifies the role, in singularity resolution,  of the three `exotic' 
structures in this LQC model, namely: curvature corrections, inverse triad definitions 
and the `polymer' nature of the kinematic representation.
We also critically examine certain technical assumptions made by APS in their analysis of 
WDW semiclassical states
and  point out some problems stemming from the infrared behaviour of their
wave functions.

\end{abstract}

\section{Introduction}

In recent years Loop Quantum Gravity (LQG) techniques have been applied to
quantize the space of homogeneous and isotropic configurations of the gravitational field
\cite{martinreview} and there is growing evidence that in the resulting quantum cosmology
(known as Loop Quantum Cosmology or LQC) the big bang singularity is resolved.
Here we focus on the LQC of the spatially flat isotropic model coupled to a homogeneous massless scalar field.
This model was studied in great detail in the LQC context 
by Ashtekar, Pawlowski and Singh (APS)\cite{aps1,aps2}.
APS consider evolution 
from a classical epoch back towards the singularity and show that quantum effects result in a `bounce' 
which occurs before the universe gets to zero size. This is the sense in which the singularity is resolved in
their work. 

APS compare the LQC quantization to that of the more conventional Wheeler DeWitt theory. In the framework of the 
Wheeler DeWitt theory they find that the singularity persists i.e. the universe reaches zero size at which point 
physical quantities of interest (such as the scalar field density) diverge. It follows that LQC is sufficiently
different from conventional quantization schemes, the differences being responsible for singularity resolution.
The three key `exotic' features of LQC in the spatially flat model are as follows:\\

\noindent
(i){\em Discreteness of spatial geometry}:
The quantum kinematics is based on an exotic representation (which is the counterpart of the 
representation used in LQG and) which endows the scale factor operator with a discrete spectrum in contrast to the
continuous spectrum obtained in the Wheeler DeWitt case wherein the scale factor operator acts by multiplication,
exactly like the position operator in particle quantum mechanics.\\

\noindent
(ii){\em  Curvature corrections}:
Analogous to the holonomy operators of LQG, the basic operator of LQC is the exponential of the 
Ashtekar- Barbero connection \cite{aabarbero} (in the flat model
under consideration, this connection is just the extrinsic curvature of the spatial slice) 
and, as a result, the quantum
dynamics needs to be re-expressed in terms of these operators. 
The 
dynamics is generated by a constraint operator whose gravitational part depends on the extrinsic curvature.
Due to the nature of the representation, the 
extrinsic curvature is not a well defined operator and APS replace this term by appropriately defined 
approximants which depend on the LQC holonomies. 
The structure of the holonomy approximants is motivated by definitions of the Hamiltonian constraint
operator in full LQG.
The approximants agree with the classical general relativistic expression at low curvatures (i.e. in the
classical regime) but differ in the  high curvature regime in the vicinity of the classical singularity.
This leads to a correction to general relativistic  dynamics near the singularity which is absent in the
Wheeler DeWitt case wherein the extrinsic curvature is a well defined operator.\\

\noindent (iii){\em  Inverse triad definitions}:
The matter part of the constraint depends on the inverse scale factor. In analogy to Thiemann's
procedure in LQG \cite{ttinverse}
this quantity is first expressed in terms of a Poisson bracket between the `holonomy'  and the 
spatial volume  and then promoted to a quantum operator by replacing the Poisson bracket by the quantum commutator.
The eigen states of the resultant inverse scale factor operator are identical with those of the scale factor operator.
Moreover, the spectrum of the resultant inverse scale factor operator agrees with that of the 
straightforward inverse  for large eigenvalues of the scale factor but differs when these eigenvalues are small,
the eigen value of the former being bounded (and typically vanishing) when the scale factor eigen value vanishes
\cite{martin,aps1,aps2}.

\vspace{3mm}

The primary aim of this work is to clarify the role of the above features in singularity
resolution in the isotropic model under consideration. 
Since we are interested in LQC, all our constructions will be based on (i).
APS have already argued persuasively (see \cite{apsrobust}) that in this model, the quantum bounce occurs
primarily due to (ii) rather than (iii). In this work we use the techniques introduced in
Reference \cite{aps2} to construct an LQC quantization {\em which does not contain feature (ii)}.
We are able to do this both with and without (iii). 
Our results are as follows.

As in \cite{aps1,aps2} the quantization admits the interpretation of the scalar field as a clock.
If one does not introduce feature (iii) i.e. if we use the straightforward scale factor operator inverse defined directly
through the spectral decomposition of the scale factor operator with the replacement of each eigen value of the 
latter by its inverse, we  obtain a representation of the physical degrees of freedom which is 
equivalent to the standard Wheeler DeWitt one. To reiterate, 
despite 
the profound differences between the quantum kinematics of LQC (on which our constructions are based)
and that of the Wheeler DeWitt (WDW) framework, the {\em physical} Hilbert space representations are identical.

More interestingly, if we retain (iii) while suppressing (ii), we obtain (modulo some hitherto un- noticed technicalities 
which are relevant to WDW part of the APS work), well defined, regular evolution through an epoch where
the size of the universe vanishes. Thus, as anticipated by APS the quantum bounce disappears due to the inavailability
of feature (ii). Nevertheless, the singularity of the classical theory is resolved in that there is regular
well defined evolution through the classically singular geometry. The physical observable of interest, namely the
scalar field energy density, is always finite even in the classically singular region due to feature (iii). In this sense
the singularity is still resolved even though there is no bounce. The details of the dynamics near the classically
singular epoch are tied to the particular choice of Thiemann like procedure used to define the inverse scale factor
operator.  

The technicalities mentioned in the beginning of the previous paragraph are related to the issue of the  validity of certain
approximations used by APS to evaluate semiclassical behaviour for the Wheeler 
DeWitt quantization of the model. Recall that the 
scale factor operator acts by multiplication so that the wave function is a function of the scale factor . 
It turns out that  the 
terms 
which APS neglect in their proposed  semiclassical wave function  significantly alter the behaviour of the 
wave function at large values of the logarithm of the scale factor. As a result, the Dirac observable corresponding to 
the scale factor operator at fixed value of the scalar field 
does not have the APS wave function in its domain.
However, (the mean value and fluctuation of) the operator corresponding to the logarithm of the scale factor   
at fixed `time' (as measured by the scalar field) is well defined for this state. 
The secondary aim of this work is to point out the existence of these technicalities. The detailed
calculations will be presented in a subsequent paper \cite{meinprep}.

The layout of this paper is as follows. We provide a brief review of the model and its quantum kinematics
in section 2. While several of the LQC constructions (see the references in \cite{martin})
were already standard prior to the seminal APS work, 
for convenience as well as to take advantage of some key APS insights
we refer the reader to References \cite{aps1,aps2} for further details; indeed we shall lean heavily on those
papers. In section 3, we show how to construct an LQC quantization without the introduction of curvature corrections.
We use key ideas from the APS work \cite{aps2}. 
In section 4, we construct the physical state space appropriate to the absence of (ii), using group averaging techniques. 
In section 5 we switch 
off both (ii) and (iii) and show that the resultant physical Hilbert space representation 
is identical to the Wheeler DeWitt one. 
We also point out the technicalities concerned with semiclassical analyis mentioned above.
In section 6, we construct the representation obtained by 
switching  off (ii) but retaining  (iii)
for the theory
and show that the singularity is resolved.
While we postpone an analysis of semiclassical states to a subsequent paper \cite{meinprep} (wherein we 
also fill the some of the lacunae in the semiclassical Wheeler DeWitt analysis of APS), our results
in \cite{meinprep} support our statement of singularity resolution. 

In section 7 we comment on the freedom in 
defining the inverse scale factor operator using Thiemann like procedures and argue that, given the interpretation
of the operator, it is reasonable to incorporate a dependence on the fiducial cell size (see \cite{aps1,aps2})
so that the statement of singularity resolution is independent of the choice of fiducial cell.
Section 8 contains a summary of our results and discusses open issues.

\section{Brief review of classical theory and quantum kinematics}

We provide a brief review of the classical Hamiltonian description of the model and its LQC quantum kinematics.
We refer the reader to References \cite{aps1,aps2} for details. Our notation and conventions agree with those of
\cite{aps1,aps2}.

\subsection{Classical Hamiltonian description}
Since the spatial slice is non- compact and the fields are homogeneous, integrals over the slice diverge necessitating
the choice of an elementary cell ${\cal V}$ in the spatial slice which serves as the domain of integration.
Fix a fiducial metric ${}^0q_{ab}$, a set of co-triads, ${}^o\omega_a^i$ and triads ${}^0e^a_i$
which are compatible with, and orthonormal with respect to fiducial metric, and let $V_0$ be the volume of the 
fiducial cell with respect to the fiducial metric. The gravitational phase space variables
are the connection $A_a^i$ and the densitized triad $E^a_i$, and are parametrised as 
\begin{equation}
A_a^i= c V_0^{-\frac{1}{3}}{}^0\omega_a^i\;\; E^a_i= pV_0^{-\frac{2}{3}}{}^0e^a_i
\end{equation}
with the symplectic structure
\begin{equation}
\{c , p \} = \frac{8\pi G\gamma}{3}
\end{equation}
where $\gamma$ is the Barbero- Immirzi parameter and $G$ is Newton's constant. 
It is easy to see that the  volume $v$ of the elementary cell in the {\em physical} metric defined by $E^a_i$ is 
\begin{equation}
V:= |p|^{\frac{3}{2}}
\label{defv}
\end{equation}
The massless scalar field $\phi$ and
its conjugate momentum $p_{\phi}$ have Poisson bracket $\{\phi , p_{\phi}\}= 1$. The diffeomorphism and Gauss Law
constraints vanish identically and the Hamiltonian constraint is 
\begin{equation}
C = -\frac{6}{\gamma^2}c^2|p|^{\frac{1}{2}} + 8\pi G \frac{p_{\phi}^2}{|p|^{\frac{3}{2}}}.
\label{hamcon}
\end{equation}
There are 4 degrees of freedom and a single constraint so that there are 2 true degrees of freedom
which indicates the necessity of a choice of two independent Dirac observables.


Let ${\vec P}= (c, p,\phi , p_{\phi})$ denote a point on the constraint surface. 
Since
$\phi =\phi_0=$ constant is a good gauge fixing, each gauge orbit can be labelled by its intersection with 
this gauge fixing slice in phase space. Let the gauge orbit through ${\vec P}$ intersect $\phi = \phi_0$ at 
${\vec P}_{\phi_0}({\vec P})= (c|_{\phi_0}, p|_{\phi_0}, \phi_0, p_{\phi}|_{\phi_0} )$. Let $f({\vec P})$
be any function on the constraint surface. Then $f({\vec P}_{\phi_0}({\vec P})):= f_{\phi_0}({\vec P})$ is
gauge invariant.
By changing $\phi_0$ we obtain a 1 parameter family of Dirac observables. These can be interpreted as 
describing the evolution of $f_{\phi_0}$ if we identify $\phi_0$ with a choice of time.

Setting $f:= p_{\phi}$, we have that $p_{\phi}|_{\phi_0}$ is a Dirac observable. Since $\{ p_{\phi}, C\} =0$, 
we have that $p_{\phi}|_{\phi_0}= p_{\phi}$. APS choose $p_{\phi}, p|_{\phi_0}$ as Dirac observables.
It cane be checked that $p|_{\phi_0}$ satisfies the equation:
\begin{equation}
\frac{dp|_{\phi_0}}{d\phi_0} = \pm \sqrt{16\pi G/3}p|_{\phi_0} . 
\label{classevoltn}
\end{equation}
The $\pm$ signs correspond to the expanding and contracting branches. For the expanding branch
equation (\ref{classevoltn}) implies that starting from some non- vanishing $p_{\phi_0=\phi_*}$ at time $\phi_*$,
and evolving backwards we have 
\begin{equation}
p|_{\phi_0\rightarrow -\infty} \rightarrow 0
\label{classsing}
\end{equation}
at which point 
the size of the universe goes to zero and the matter density $\frac{p_{\phi}}{V_{\phi_0}}$ diverges (here 
we have used the notation above and set $f= V$ to define $V_{\phi_0}$).
This is the Big Bang singularity and every expanding classical solution originates from it.

We shall find it convenient to choose $p_{\phi}, x_{\phi_0}$ as Dirac observables where $x$ is an 
appropriately defined function of $p$ (see equation (\ref{defhatxdirac})).

%

\subsection{LQC quantum kinematics}
The basic operators of LQC in the gravity sector 
are ${\hat e^{i\lambda c}},\beta \in R$ and ${\hat p}$. Their action on eigen states of
${\hat p}$ is
\begin{equation}
{\hat e^{i\lambda c}}|\mu\rangle= |\mu +\lambda\rangle \;\;\; {\hat p}|\mu\rangle= \frac{8\pi\gamma l_P^2}{6} \mu |\mu\rangle
\end{equation}
where $l_P^2= G\hbar$ and $\mu\in R$. The Hilbert space ${\cal H}^{grav}_{kin}$ 
is spanned by eigen states of $\mu$ and the inner product is
defined through
\begin{equation}
\langle \mu_1 |\mu_2\rangle = \delta_{\mu_1, \mu_2}
\end{equation}
where  $\delta_{\mu_1, \mu_2} = 1$ if $\mu_1= \mu_2$ and vanishes otherwise. 
While ${\hat e^{i\lambda c}}$ are unitary operators, the above inner product does not endow them with enough continuity
in $\lambda$ for ${\hat c}$ to be defined as an operator on ${\cal H}^{grav}_{kin}$.
The matter operators are represented in the standard $L^2(R, d\phi)$ representation wherein ${\hat \phi}$ acts by 
multiplication and ${\hat p_{\phi}}:= \frac{\hbar}{i}\frac{d}{d\phi}$. 

The kinematic Hilbert space ${\cal H}_{kin}$ for the model is just the product space 
${\cal H}^{grav}_{kin}\otimes L^2(R, d\phi)$.

\section{Quantization without curvature corrections}
Since ${\hat c}$ is not defineable on ${\cal H}_{kin}$, APS replace 
${\hat c}$ in the Hamiltonian constraint by $\frac{{\hat e^{i\lambda c}}-{\hat e^{i\lambda c}}}{2i\lambda}$ for small 
(but necessarily non-vanishing) $\lambda$. Since $\lambda \neq 0$, this amounts to the addition of the higher 
order curvature corrections  (to the general relativistic expression of the  Hamiltonian
constraint) alluded to in (ii) of section 1. While $\lambda$ was chosen to be a fixed number in \cite{aps1}, this was
improved upon from a physical standpoint in \cite{aps2} wherein $\lambda$ was allowed to be operator valued and 
dependent on ${\hat p}$.  
The construction and well defined action  of the 
operator ${\hat e^{i\lambda c}}$ with $\lambda$ being an operator valued function of ${\hat p}$, 
was one of the key insights of APS \cite{aps2}.
We shall use this key insight of APS in conjunction with the group averaging technique \cite{alm^2t}
to construct a quantization of the model without curvature corrections.

First, note that the Hamiltonian constraint $C$ may be written as 
\begin{equation}
C= -C_+C_-
\end{equation} 
where
\begin{equation}
C{\pm}= -\sqrt{\frac{6}{\gamma^2}}c|p|^{\frac{1}{4}} \pm \sqrt{ {8\pi G }}\frac{p_{\phi}}{|p|^{\frac{3}{4}}}
\end{equation}
so that the vanishing of $C$ is equivalent to that of $C_+$ or $C_-$ or both.
We shall construct the physical Hilbert space as the union of the kernels of ${\hat C}_{\pm}$. However,
since it is not possible to define ${\hat C}_{\pm}$  as operators, 
we shall first define their exponentials, $e^{i\lambda C_{\pm}}$
as unitary operators and then find the physical state space by group averaging the action of these unitary operators.

We shall work in the $\mu$ representation. We start with the following heuristics.
Following APS we define ${\hat C}_{\pm}$ by  ${\hat c}= 2i \frac{d}{d\mu}$ and ${\hat p}= \frac{8\pi\gamma l_P^2}{6}\mu$
(this is heuristic since ${\hat c}$ is not a well defined operator on ${\cal H}_{kin}$).
In anticipation of a Thiemann like definition of the operator ${\hat {|p|^{-\frac{3}{2}}}}$ (which we assume is 
diagonal in the $\mu$ representation as in \cite{aps1,aps2}), we set
\begin{equation}
{\hat {|p|^{-\frac{3}{2}}}} = ({\frac{4\pi}{3} 
\gamma l_P^2})^{-\frac{3}{2}}B(\mu ) 
\end{equation}

\vspace{3mm}

It is useful to define
\begin{equation}
{\tilde C}_{\pm}:= (4 (\frac{3\pi l_P^2}{\gamma^3})^{\frac{1}{4}})^{-1} C_{\pm}.
\label{tildecpm}
\end{equation}
Motivated by the considerations in the previous paragraph, we define the action of
the operators $\hat {e^{i\alpha {\tilde C}_{\pm}}}$ in the $\mu$ representation by  
\begin{equation}
{\hat {e^{i\alpha {\tilde C}_{\pm}}}}
= \exp \alpha (|\mu |^{\frac{1}{4}}\frac{d}{d\mu} \pm i \frac{B^{1/2}(\mu ){\hat p}_{\phi}}{\sqrt{\frac{16\pi G\hbar^2}{3}}}),
\end{equation}
where $\alpha \in R $.
Again, following the ideas of APS we define
\begin{equation}
l:= \frac{4}{3} {\rm sgn}(\mu ) |\mu |^{\frac{3}{4}}.
\label{defl}
\end{equation}
Note that $l$ is an invertible function of $\mu$. From (\ref{defl}) 
it follows that 
\begin{equation}
|\mu |^{\frac{1}{4}}\frac{d}{d\mu} = \frac{d}{dl}.
\label{ddl}
\end{equation}
With this change of variables it follows that 
\begin{eqnarray}
{\hat {e^{i\alpha {\tilde C}_{\pm}}}}
&=& \exp \alpha (\frac{d}{dl} \pm i \frac{B^{1/2}(\mu){\hat p}_{\phi}}{\sqrt{\frac{16 \pi G\hbar^2}{3}}})
\\
&=& e^{\mp  i \frac{x(l){\hat p}_{\phi}}{\sqrt{\frac{16 \pi G\hbar^2}{3}}}}
e^{\alpha \frac{d}{dl}} 
e^{\pm  i \frac{x(l){\hat p}_{\phi}}{\sqrt{\frac{16 \pi G\hbar^2}{3}}}}.
\label{ecpm}
\end{eqnarray}
Here $x(l)$ is defined as
\begin{equation}
x(l) = \int_{L}^{l}B^{1/2}(\mu ({\bar l})) d{\bar l}
\label{defx}
\end{equation}
with some (fixed) choice of lower limit of integration $L$.

Since $\mu$ is an invertible function of $l$,
it is convenient to label the eigenstates of ${\hat p}$ by $l$ rather than $\mu$ so that 
\begin{eqnarray}
{\hat p}|l\rangle &:=& \frac{8\pi \gamma l_P^2}{6}
\mu (l)|l\rangle = {\rm sgn}(l) \frac{3}{4} |l|^{\frac{4}{3}} |l\rangle
\label{defketl}\\
\langle l_1 | l_2\rangle &=& \delta_{l_1,l_2} 
\label{lip}
\end{eqnarray}
Thus any state $|\psi\rangle \in {\cal H}_{kin}$ can be written as 
\begin{equation}
|\psi\rangle = \sum_l \int d\phi \psi (l,\phi ) |l\rangle \otimes |\phi \rangle .
\label{defpsilphi}
\end{equation} 
In this $l$- representation, it follows from the inner product (\ref{lip}) and the inner product on 
$L^2( R,d\phi )$ that the inner product 
between two states $|\psi_1\rangle ,  |\psi_2\rangle $ is
given by 
\begin{equation}
\langle \psi_1 | \psi_2 \rangle = \sum_l \int \psi_1^* (l, \phi ) \psi_2 (l, \phi )
\label{ippsilphi}
\end{equation}

From (\ref{ecpm}) it follows that the action of the exponentiated constraint operators
$\hat {e^{i\alpha {\tilde C}_{\pm}}}$ on the state $|\Psi \rangle$ defined by equation (\ref{defpsilphi}) is defined
to be 
\begin{equation}
{\hat {e^{i\alpha {\tilde C}_{\pm}}}}|\psi \rangle
= \sum_l \int d\phi 
\psi (l, \phi \pm \beta_0(x(l) - x(l- \alpha ))) 
|l-\alpha \rangle \otimes |\phi \rangle
\label{lecpm}
\end{equation}
where we have defined $\beta_0$ by 
\begin{equation}
\beta_0 := (\sqrt{\frac{16\pi G}{3}})^{-1}
\label{defbeta0}
\end{equation}
It can be checked that this action is unitary in the inner product (\ref{ippsilphi}) on ${\cal H}_{kin}$.

Thus, motivated by the ideas of APS \cite{aps2}, we have constructed the well defined unitary operators 
$\hat {e^{i\alpha {\tilde C}_{\pm}}}$ on ${\cal H}_{kin}$ without recourse to any curvature
corrections. In the next section we contruct the kernel of the constraints 
$C_{\pm}$ (or, equivalently 
${\tilde C}_{\pm}$ (see equation (\ref{tildecpm})) by group averaging the action of the operators
defined in equation (\ref{lecpm}).  As mentioned above, the kernel of the Hamiltonian constraint will be 
identified with the union of the kernels of $C_+$ and $C_-$. 


\section{Group Averaging and the Physical Hilbert Space}

We start with a brief review of the group averaging technique.
Only gauge invariant states are physical so that physical states $\Psi$  must satisfy the condition 
${\hat U}(g )\Psi = \Psi, \; \forall g$ where 
${\hat U}(g)$ is the unitary operator which implements the finite gauge transformation denoted by $g$. 
A formal solution to this condition is to fix some
$|\psi\rangle \in {\cal H}_{kin}$ and set
$\Psi = \sum |\psi^{\prime}\rangle$ where the sum is over all distinct $|\psi^{\prime}\rangle$  which are
gauge related to $|\psi\rangle$.  A mathematically precise implementation of this idea places the gauge invariant
states in the dual representation (corresponding to a formal sum over bras rather than kets) and goes by the name of
Group Averaging. The ``Group'' is that of gauge transformations and the ``Averaging'' corresponds to
the construction of a gauge invariant state from a kinematical one by giving meaning to the formal sum over
gauge related states. Specifically (for details see Reference \cite{alm^2t}), the physical Hilbert space
can be constructed if there exists an anti-linear map $\eta$ from a dense subspace ${\cal D}$ of the kinematical
Hilbert space ${\cal H}_{kin}$,
to its algebraic dual ${\cal D}^*$, subject to certain requirements. The algebraic dual of  ${\cal D}$
is defined to be the space of linear mappings from ${\cal D}$ to the complex numbers. The requirements which $\eta$
needs to satisfy are as follows. Let $|\psi_1\rangle , |\psi_2\rangle \in {\cal D}$, let ${\hat A}$ be a (strong) 
Dirac observable of interest
and let $g$ be a gauge transformation with ${\hat U} (g )$ being its unitary implementation on ${\cal H}_{kin}$.
Let $\eta (|\psi_1\rangle )\in {\cal D}^*$ denote the image of $|\psi_1\rangle$ by $\eta$ and let 
$\eta (|\psi_1\rangle )[|\psi_2\rangle ]$ denote the
complex number obtained by the action of $\eta (|\psi_1\rangle )$ on $|\psi_2\rangle$.
Then for all $|\psi_1\rangle , |\psi_2\rangle , {\hat A},g$ we require that \\
\noindent {\bf (1)} $\eta (|\psi_1\rangle )[|\psi_2\rangle ]= \eta (|\psi_1\rangle )[{\hat U} (g )|\psi_2\rangle ] $\\
\noindent {\bf (2)} $\eta (|\psi_1\rangle )[|\psi_2\rangle ] =(\eta (|\psi_2\rangle )[|\psi_1\rangle ])^*$, 
$\eta (|\psi_1\rangle )[|\psi_1\rangle ] \geq 0$. \\
\noindent {\bf (3)} $\eta (|\psi_1\rangle )[{\hat A}|\psi_2\rangle ]= \eta ({\hat A}^{\dagger}|\psi_1\rangle )
[|\psi_2\rangle ] $.\\

It turns out that typically (, and in the case of interest here,) we may indeed write 
$\eta (|\psi  \rangle) = \sum \langle \psi^{\prime}|$ where the sum is over all distinct $\langle\psi^{\prime}|$  
which are
gauge related to $\langle \psi |$. As we shall see only a finite number of terms in the sum have a non- vanishing
kinematical inner product with any  state in ${\cal D}$ so that  $\eta (|\psi  \rangle)$ (with its action defined on 
states in ${\cal D}$ in the obvious, natural way suggested by its representation by the sum above) indeed lies in the 
algebraic dual space. An inner product on the space $\eta ({\cal D} )$ can be defined through
\begin{equation}
<\eta (|\psi_1\rangle), \eta (|\psi_2\rangle)>=
\eta (|\psi_1\rangle) [|\psi_2\rangle ]. 
\label{physip}
\end{equation}
The requirements ${\bf (1)},{\bf (2)}$ ensure that the right hand side of the above equation defines a 
positive, hermitian inner product. The completion of $\eta ({\cal D})$ in this inner product is 
the physical Hilbert space. It can be checked that the condition ${\bf (3)}$ ensures that 
the above inner product automatically implements  the adjointness conditions on the Dirac observables (which act by
dual action on ${\cal D}^{*}$)
\footnote{Given $\Psi  \in {\cal D}^{*}$, $|\psi\rangle  \in {\cal D}$ and ${\hat A}$ such that 
${\hat A}^{\dagger}|\psi\rangle \in {\cal D}$,
define ${\hat A}\Psi$ through 
${\hat A}\Psi  [|\psi\rangle  ]:= \Psi [{\hat A}^{\dagger}|\psi \rangle ]$. This is the dual action.
\label{dualaction}
}
if these conditions are  implemented on ${\cal H}_{kin}$.

As mentioned in section 3 our strategy is to construct the kernel of the Hamiltonian constraint as the 
union of the $+$ and $-$- sector kernels. Accordingly,
in section 4.1 we define the group averaging maps $\eta^{\pm}$ corresponding to the group averaging 
with respect to ${\hat {e^{-i\alpha {\tilde C}_{\pm}}}}$ and construct the corresponding physical Hilbert spaces 
${\cal H}_{phys}^{\pm}$.

In section 4.2 we construct the Dirac observables of the theory.
In section 4.3 we identify the positive and negative frequency
eigen states of the Dirac observable ${\hat p_{\phi}}$ within each sector ${\cal H}_{phys}^{\pm}$.
Denote the positive frequency subspaces  by ${\cal H}_{+phys}^{\pm}$.
As in the APS work, we shall restrict attention to these positive frequency subspaces.
The space of positive frequency physical states is the union of 
${\cal H}_{+phys}^{+}$ and ${\cal H}_{+phys}^{-}$ 
While group averaging automatically provides the inner product between states within each sector
${\cal H}_{phys}^{\pm}$, it does not specify the inner product between a state in 
${\cal H}_{phys}^{+}$ and one in  ${\cal H}_{phys}^{-}$. In section 4.4 we show that the positive frequency subspaces 
of 
${\cal H}_{phys}^{+}$ and ${\cal H}_{phys}^{-}$ must be mutually orthogonal. Thus the 
poistive frequency physical Hilbert space of the model, ${\cal H}_{+phys}$, 
is the union of its two mutually orthogonal positive frequency subspaces 
${\cal H}_{+phys}^{+}$ and ${\cal H}_{+phys}^{-}$.
\footnote{Similar arguments show that the negative frequency subspaces, ${\cal H}_{-phys}^{\pm}$
 of ${\cal H}_{phys}^{\pm}$ are
mutually orthogonal. Their union, 
${\cal H}_{-phys}$, is the negative frequency physical Hilbert space of the model. Note that 
if we wish to work with  both positive and negative frequency states, such states must be mutually orthogonal
to ensure hermiticiy of the Dirac observable ${\hat p}_{\phi}$.}
Finally, in section 4.5 we show that the representation on ${\cal H}_{+phys}$ is isomorphic
to an $L^2 (R, dx )$ representation.

\subsection{Construction of ${\cal H}_{phys}^{\pm}$ by group averaging.}
Consider states of the form 
\begin{equation}
|\psi\rangle = \int d\phi \psi (\phi ) |l\rangle \otimes |\phi\rangle
\label{psil=l}
\end{equation}
where $\psi (\phi )$ is smooth and normalizable in $L^2(R, d \phi )$. Clearly 
the finite span of 
such states defines  a dense set
${\cal D}\subset {\cal H}_{kin}$. Let ${\cal D}^*$ be its algebraic dual. Define the group averaging 
maps $\eta^{\pm}$ from ${\cal D}$ to ${\cal D}^*$ through
\begin{equation}
\eta^{\pm}(|\psi\rangle ) := \sum_{\alpha} \langle \psi | {\hat {e^{-i\alpha {\tilde C}_{\pm}}}}
\label{defetapm}
\end{equation}
where the formal sum is over all $\alpha \in R$ and the right hand side is interpreted as an element
of ${\cal D}^*$ in the usual way \cite{alm^2t}. For ease of notation in what follows, we set $l=l_0$
in (\ref{psil=l}) so that 
\begin{equation}
|\psi\rangle := \int d\phi \psi (\phi ) |l_0\rangle \otimes |\phi\rangle .
\label{psil=l0}
\end{equation}
Then from (\ref{lecpm}), (\ref{defetapm}), (\ref{psil=l0}) the action of $\eta_{\pm}$ on $|\psi\rangle$
evaluates to
\begin{eqnarray}
\eta^{\pm}(|\psi\rangle )&=&
\sum_{\alpha} \int d\phi \psi^* (\phi \pm \beta_0(x(l_0) - x(l_0- \alpha ))) 
\langle l_0-\alpha | \otimes \langle \phi | 
\label{etapmexplicit1}\\
&=&
\sum_{l} \int d\phi \psi^* (\phi \pm \beta_0(x(l_0) - x(l))) 
\langle l| \otimes \langle \phi | .
\label{etapmexplicit2}
\end{eqnarray}
where, in the second line, the sum is over all $l \in R$.

It can be checked that, for any $\lambda \in R$,
\begin{equation}
\eta^{\pm} (\hat {e^{i\lambda {\tilde C}^{\pm}}}|\psi \rangle ) = \eta^{\pm} (|\psi \rangle ) .
\label{etaecpm}
\end{equation}
Since 
the orbit of $\langle l_0| $ under the averaging procedure is $\{ \langle l | , l \in R \} $,
\footnote{ This is not strictly correct. As we shall see in section 6, this depends on the 
behaviour of $x(l)$. The attendant subtelities will be dealt with in section 6.
\label{footnote1}}
equation 
(\ref{etaecpm}) implies that we may, without loss of generality, generate a basis for the physical
state space by averaging over states of the form (\ref{psil=l0}) with $l_0$ fixed once and for all.
Next, consider the states $|\psi_1\rangle , |\psi_2\rangle$ of the form 
(\ref{etaecpm}) with $\psi (\phi ) = \psi_1 (\phi ), \psi_2 (\phi )$ respectively.
The inner product between the corresponding physical states obtained by group averaging, 
$(\eta^{\pm}(|\psi_1\rangle ), \eta^{\pm}(|\psi_2\rangle ))$, is defined as
$(\eta^{\pm}(|\psi_1\rangle ), \eta^{\pm}(|\psi_2\rangle )):= \eta^{\pm}(|\psi_2\rangle )[|\psi_1\rangle]$
where the $\eta_{\pm}(|\psi_2\rangle )[|\psi_1\rangle]$ denotes the natural action of elements of 
${\cal D}^*$ on elements of ${\cal D}$ \cite{alm^2t}. From (\ref{etapmexplicit2}) this evaluates to
\begin{equation}
(\eta^{\pm}(|\psi_1\rangle ), \eta^{\pm}(|\psi_2\rangle ))= \int d\phi \psi_2^*(\phi )\psi_1 (\phi )
\label{iphphyspm}
\end{equation}
which is clearly positive definite.
\footnote{One may also attempt to define the averaging map with respect to the averaging measure $\int d\alpha $.
If one does this with the same choice of ${\cal D}$ as above, one finds that all of ${\cal D}$ is in the 
kernel of the group averaging maps or, equivalently, the physical inner product 
is completely degenerate.}

The completion of $\eta_{\pm}({\cal D})$ in the inner product (\ref{iphphyspm}) yields the physical Hilbert 
spaces
${\cal H}_{phys}^{\pm}$.

\subsection{The Dirac Observables}
Recall from section 2 that ${\hat p}_{\phi}$ is one of our Dirac observables. It is straightforward to check that 
${\hat p}_{\phi}$ commutes with $\hat{ e^{i\alpha{\tilde C}_{\pm}}}$ as well as with the averaging 
maps $\eta_{\pm}$. Hence \cite{alm^2t} it's kinematic self adjontness translates to self
adjointness on the physical Hilbert spaces ${\cal H}_{phys}^{\pm}$.

Next recall from (\ref{defl}), (\ref{defx}) that $x:= x(l)= x(l(\mu ))$. Since the eigen values of
${\hat p}$ are $\frac{8\pi \gamma l_P^2}{6} \mu$, we define 
\begin{equation}
{\hat x} = x(l(\frac{\hat p}{\frac{8\pi \gamma l_P^2}{6}})) =: f ({\hat p}).
\label{defhatxdirac}
\end{equation}
Using the notation of section 2.1, we choose $f(\hat{p|_{\phi_0}}) = {\hat x}_{\phi_0}$ as our second Dirac observable.

In order to represent this operator, it is useful to examine its classical correspondent, 
$x_{\phi_0}$.  To evaluate $x_{\phi_0}$ at any point on the constraint surface we first map the point in 
question via a gauge transformation to its gauge related image on the gauge fixing slice $\phi =\phi_0$
and then evaluate the function $x$ there. The constraint surface splits (modulo a set of measure zero) 
into $+$ and $-$ sectors defined
by $C_+=0, C_-\neq 0$ and  $C_-=0, C_+\neq 0$ respectively. Denote the restriction of the function 
$x_{\phi_0}$ to the $\pm$ sector by $x^{\pm}_{\phi_0}$. Correspondingly, in quantum theory, we define the action 
of ${\hat x}^{\pm}_{\phi_0}$ on an eigen state of ${\hat \phi}$ by first mapping the state to the eigen state
with eigenvalue $\phi_0$ by an appropriate gauge transformation of the form $\hat{e^{i\alpha {\tilde C}_{\pm}}}$,
acting with ${\hat x}$ and then performing the inverse gauge transformation.

Hence the action of the operators ${\hat x}^{\pm}_{\phi_0}$ relevant to the physical Hilbert spaces 
${\cal H}_{phys}^{\pm}$ is obtained as follows. From (\ref{lecpm}),
\begin{equation}
{\hat{e^{i\alpha \tilde{C}_{\pm}}}} |l\rangle\otimes |\phi \rangle
= |l-\alpha \rangle\otimes |\phi\pm \beta_0(x(l)-x (l-\alpha )) \rangle ,
\end{equation}
where $\alpha$ is chosen so that
$\phi_0 = \phi \mp \beta_0 (x(l) - x(l - \alpha ) ) $ which implies that 
$x( l- \alpha ) = x(l) \pm (\phi_0 - \phi )$.  It follows that 
\begin{equation}
{\hat x}^{\pm}_{\phi_0} |l\rangle\otimes |\phi \rangle
= ( x(l) \pm \frac{\phi - \phi_0}{\beta_0} ) |l\rangle\otimes |\phi \rangle ,
\label{defxhatpm}
\end{equation}
so that 
\begin{eqnarray}
{\hat x}_{\phi_0}(\eta^{\pm}( |\psi \rangle )
&:= &{\hat x}^{\pm}_{\phi_0}(\eta^{\pm}( |\psi \rangle )\\
&=& \sum_l \int d\phi \langle l|\otimes \langle \phi | 
\chi^* (\phi \pm \beta_0 ( x(l_0) - x(l) ) ) ,
\label{defhatxphifinal}
\end{eqnarray}
where $|\psi \rangle$ is given by equation (\ref{psil=l0}) and 
\begin{eqnarray}
\chi (\phi \pm \beta_0 ( x(l_0) - x(l) ) )
&:=& \mp \beta_0^{-1} \{ \phi \pm \beta_0 (x(l_0)- x(l) )  -  (\phi_0 \pm \beta_0 x(l_0))\} \nonumber \\
&&\;\;\;\;\psi (\phi\pm \beta_0 ( x(l_0) - x(l) ) )
\label{defchi}
\end{eqnarray}

It can be verified that $[{\hat x}_{\phi_0}, {\hat{e^{i\lambda \tilde{C}_{\pm}}}}]= 0, \lambda \in R$,
that $[{\hat x}_{\phi_0}, \eta^{\pm} ]=0$ and that 
${\hat x}_{\phi_0}$ is a self adjoint operator on ${\cal H}^{\pm}_{phys}$.

\subsection{Eigen functions of ${\hat p_{\phi}}$ in ${\cal H}_{phys}^{\pm}$.}
Since ${\hat p}_{\phi}$ commutes with the averaging maps $\eta_{\pm}$ it follows that 
group averages of kinematic eigen states of ${\hat p_{\phi}}$ are eigen states of 
${\hat p_{\phi}}$ with unchanged eigen values.
The (positive and negative) frequency kinematic  eigen states of ${\hat p_{\phi}}$ are
\begin{equation}
\psi_{\pm \omega}(\phi ) = e^{\pm i\omega \phi} \;\; \omega >0 .
\label{pmomegakin}
\end{equation}
We shall restrict attention to the positive frequency states.
From equation (\ref{etapmexplicit2}) these states under group averaging yield (upto an unimportant constant 
phase factor which we drop),
\begin{equation}
\eta^{\pm}( |\psi_{\omega} \rangle ):=
\sum_{l} \int d\phi e^{- i\omega \phi} e^{\pm i\beta_0\omega x(l)}
\langle l |\otimes \langle \phi | .
\label{pmomegaphys}
\end{equation}
The physical positive frequency eigenstates,  $\eta^{\pm}( |\psi_{\omega} \rangle )$, form a spanning set in 
the positive frequency physical Hilbert spaces ${\cal H}_+^{\pm}$. This follows from the fact that the 
kinematic positive frequency eigenstates,
$|\psi_{\omega}\rangle= \int d\phi e^{i\omega \phi}|l\rangle \otimes |\phi \rangle$
span the positive frequency part of the kinematic Hilbert space.

Finally, note that from (\ref{iphphyspm}) it follows that 
\begin{equation}
(\eta^{\pm}( |\psi_{\omega_1} \rangle ),\eta^{\pm}( |\psi_{\omega_2} \rangle )
= 2\pi \delta (\omega_1 , \omega_2 )
\label{ipestate}
\end{equation}
where $\delta (\omega_1 , \omega_2 )$ is the Dirac delta function.

\subsection{Mutual orthogonality of ${\cal H}_{+phys}^{+}$, ${\cal H}_{+phys}^{-}$.}

Our strategy is as follows. The phase space function, $F$, defined by 
\begin{equation}
F = - (\frac{6}{8 \pi G \gamma^2})^{1/2}c|p|
\label{deff}
\end{equation}
is a (weak) Dirac observable. We shall represent ${\hat F}$ as an operator on ${\cal H}_{phys}^{\pm}$
and show that the positive frequency eigen functions $\eta^{\pm}( |\psi_{\omega} \rangle )$ are also
eigen functions of ${\hat F}$ with eigen values $\mp \omega$. In order that ${\hat F}$ be
represented as a self adjoint operator, its eigen spaces with distinct eigenvalues must be orthogonal. The mutual 
orthogonality of ${\cal H}_{+phys}^{\pm}$ follows.

From considerations similar to those of section 3, we represent the operator 
$\hat{e^{\frac{i\beta_0\lambda}{\hbar}F}}$ on states $|\psi \rangle$ of the form (\ref{psil=l0}) by 
\begin{equation}
{\hat {e^{\frac{i\beta_0\lambda}{ \hbar}F}}} |\psi \rangle
= \int d\phi \psi (\phi ) |x^{-1}[x(l_0)-\lambda\rangle \otimes |\psi \rangle .
\label{defefhat}
\end{equation}
This follows from the (heuristic) following choice of representation for ${\hat F}$ in the $l$- representation:
\begin{equation}
{\hat F}= - (\frac{6}{8 \pi G \gamma^2})^{1/2}{\hat {c|p|^{1/4}}} \frac{1}{{\hat |p|^{-3/4}}}
= -i\hbar (\beta_0)^{-1} B^{-1/2}(\mu )\frac{d}{dl}= -i\hbar (\beta_0)^{-1}\frac{d}{dx(l)} .
\label{fhatheur}
\end{equation}
Again, we shall ignore the subtelities indicated in Footnote \ref{footnote1} and assume that $x(l)$ is an invertible
function of $l$. We shall clarify these subtelities in sections 5 and 6.

It is straightforward to check that, under the assumption of invertibility of $x(l)$, 
$\hat{e^{\frac{i\beta_0\lambda}{\hbar}F}}$ is a unitary operator.
Next, note that $\hat{e^{\frac{i\beta_0\lambda}{\hbar}F}}$ commutes with the averaging maps $\eta^{\pm}$.
To see this, note that equations (\ref{psil=l0},(\ref{etapmexplicit2}) and (\ref{defefhat}) imply that 
\begin{equation}
\eta^{\pm}({\hat{ e^{\frac{i\beta_0\lambda}{ \hbar}F}}}|\psi \rangle )
=
\sum_{l} \int d\phi \psi^* (\phi \pm \beta_0(x(l_0) - (x(l)+\lambda ))) 
\langle l| \otimes \langle \phi | .
\label{etapmef}
\end{equation}
On the other hand, we have that 
\begin{eqnarray}
{\hat{ e^{\frac{i\beta_0\lambda}{\hbar}F}}}\eta^{\pm}(|\psi \rangle )
=
\sum_{l}& \int d\phi \psi^* (\phi \pm \beta_0(x(l_0) - x(l)) 
\langle l| \otimes \langle \phi |{\hat{ e^{-\frac{i\lambda}{\beta_0 \hbar}F}}}&\nonumber\\
=
\sum_{l}& \int d\phi \psi^* (\phi \pm \beta_0(x(l_0) - x(l)) 
\langle x^{-1}[x(l)-\lambda ]| \otimes \langle \phi |& \\
=
\sum_{{\bar l}}& \int d\phi \psi^* (\phi \pm \beta_0(x(l_0) - (x({\bar l})+\lambda ))) 
\langle {\bar l}| \otimes \langle \phi |  ,&
\label{efetapm}
\end{eqnarray}
where in the last line we have set ${\bar l} = x^{-1}(x(l) -\lambda)$. The assumed invertibility of $x (l)$
ensures that (\ref{efetapm}) is the same as the right hand side of (\ref{etapmef}). Thus,
$\hat{e^{\frac{i\beta_0\lambda}{\hbar}F}}$ commutes with the averaging maps and defines an operator on the 
physical Hilbert spaces ${\cal H}^{\pm}$.

Next, note from (\ref{etapmef}) that
\begin{equation}
\lim_{\lambda \rightarrow 0} \frac{{\hat{ e^{\frac{i(\beta_0\lambda }{\hbar}F}}}- 1}{\lambda}
\eta^{\pm}(|\psi \rangle ) = 
\sum_{l} \int d\phi [\frac{d}{dx(l)}\psi^* (\phi \pm \beta_0(x(l_0) - x(l))) ]
\langle l| \otimes \langle \phi |
\label{limepsilon}
\end{equation}
Clearly, this allows us to define the action of the operator ${\hat F}$ on physical states through the action
\begin{equation}
{\hat F}\eta^{\pm}(|\psi \rangle ) =
\sum_{l} \int [(i\hbar (\beta_0)^{-1}   \frac{d}{dx(l)})\psi^* (\phi \pm \beta_0(x(l_0) - x(l))) ]
\langle l| \otimes \langle \phi |
\label{deffhat}
\end{equation}
With this definition of ${\hat F}$ it is easy to see that 
\begin{equation}
{\hat F}\eta^{\pm}(|\psi_{\omega} \rangle )
= \mp \omega \eta^{\pm}(|\psi_{\omega} \rangle ) .
\end{equation}
Hermiticity then requires the mutual orthogonality of ${\cal H}_{+phys}^+$ and ${\cal H}_{+phys}^-$.

\subsection{Isomorphism with an $L^2(R, dx )$ representation.}
The representation we have constructed through group averaging is an anti- representation due to the dual
action of operators on ${\cal D}^*$. Hence its conjugate representation is a true representation.
The state conjugate to that in (\ref{etapmexplicit2}) is characterised by the wave function
$\psi (\phi \pm \beta_0 (x(l_0)- x(l)))$. Henceforth, with no loss of generality, we shall set 
$l_0= L$ so that $x(l_0 )=0$ (see (\ref{defx})). Thus, wave functions in the conjugate Hilbert spaces
${\cal H}^{*\pm}_{phys}$ are of the form 
\begin{equation}
\psi^{\pm}(l, \phi )= \psi^{\pm}(\phi \mp \beta_0 x(l)) .
\end{equation}
Clearly, the operator ${\hat p}_{\phi}$ acts as
\begin{equation}
{\hat p}_{\phi} \psi^{\pm}(l, \phi ) = \frac{\hbar}{i}\frac{d}{d\phi} \psi^{\pm}(l, \phi ).
\label{4.5.1}
\end{equation}
The positive frequency eigenstates are 
\begin{equation}
\psi_{\omega}^{\pm}(l, \phi ) = e^{i\omega \phi} e^{\mp \beta_0 \omega x(l)}.
\label{52}
\end{equation}
For the remainder of this section we shall restrict attention to positive frequency wave functions.
From equation (\ref{defchi}) we have that 
\begin{equation}
\beta_0 {\hat x}_{\phi_0}\psi^{\pm}_{\omega} (l, \phi )
= [\mp(\phi \mp \beta_0 x(l) ) \pm \phi_0 ]\psi^{\pm}_{\omega} (l, \phi ).
\label{53}
\end{equation}
Any state in the positive frequency Hilbert space ${\cal H}^{*\pm}_+$ can be expanded as
\begin{equation}
\psi^{\pm}(l, \phi )= \int_{0}^{\infty} d\omega f_{\pm}(\omega )\psi^{\pm}_{\omega} (l, \phi ).
\label{54}
\end{equation}
It is easy to check that 
\begin{eqnarray}
\beta_0 {\hat x}_{\phi_0}\psi^{\pm} (l, \phi )
&=& \mp i \int_{0}^{\infty} d\omega \frac{df_{\pm}(\omega )}{d\omega}\psi^{\pm}_{\omega} (l, \phi )
\nonumber\\
&& \mp i( f_{\pm}(0) - f_{\pm}(\infty )) \pm \phi_0 \psi^{\pm} (l, \phi ).
\label{f0term}
\end{eqnarray}
Hence ${\hat x}_{\phi_0}$ is a well defined operator if $f_{\pm}(0)= f_{\pm}(\infty )=0$.
For arbitrary powers of ${\hat x}_{\phi_0}$ to be (densely) defined we require that $f_{\pm}(\omega )$ and
all its derivatives vanish at zero and infinity i.e.
we require that for all positive integers $n$, 
\begin{equation}
f_{\pm}(\omega ), \frac{d^n}{d\omega^n}f_{\pm} (\omega ) \rightarrow 0  {\rm as}\;  \omega \rightarrow 0, \infty .
\label{hatxphi0domain}
\end{equation}
On this dense domain, $\beta_0{\hat x}_{\phi_0}$ acts as 
\begin{equation}
\beta_0 {\hat x}_{\phi_0}\psi^{\pm} (l, \phi )
=\mp i \int_{0}^{\infty} d\omega \frac{df_{\pm}(\omega )}{d\omega}\psi^{\pm}_{\omega} (l, \phi )
\pm \phi_0 \psi^{\pm} (l, \phi ).
\label{4.5.2}
\end{equation}
Rewrite $\psi^- (l, \phi )$ as
\begin{equation}
\psi^- (l, \phi )= \int_{-\infty}^{0}d \omega f_-(- \omega )e^{i|\omega |\phi}e^{-i\omega \beta_0 x}
\end{equation}
so that  $\beta_0{\hat x}_{\phi_0}$ acts as 
\begin{equation}
\beta_0 {\hat x}_{\phi_0}\psi^{-} (l, \phi )
= -i \int_{-\infty}^0 d\omega \frac{d}{d\omega}f_-(-\omega) e^{i|\omega |\phi}e^{-i\omega \beta_0 x}
- \phi_0\psi^{-} (l, \phi ).
\label{4.5.3}
\end{equation}
The inner product on ${\cal H}_+^{*} = {\cal H}_+^{*+}\oplus {\cal H}_+^{*-}$ is defined through
\begin{equation}
(\psi_{\omega}^{\pm} , \psi_{\omega^{\prime}}^{\pm} ) =2 \pi \delta (\omega , \omega^{\prime} )\;\;
(\psi^+_{\omega}, \psi^-_{\omega^{\prime}} ) = 0 , \;\; \omega, \omega^{\prime} >0, 
\end{equation}
which yields the inner product between any $\psi_1(l,\phi ),\psi_2 (l,\phi )\in{\cal H}_{+phys}^{*}$
(in obvious notation):
\begin{equation}
(\psi_1 , \psi_2 )= 2\pi [\int_{0}^{\infty}f_{1+}^*(\omega )f_{2+}(\omega )d\omega
               + \int_{-\infty}^{0}f_{1-}^*(-\omega )f_{2-}(-\omega )d\omega ]
\label{4.5.4}
\end{equation}
where 
\begin{equation}
\psi_i (l, \phi )= \psi_i^+ (l , \phi ) \oplus  \psi_i^- (l, \phi ), \; i=1,2.
\label{4.5.5}
\end{equation}
Equations (\ref{4.5.1}),(\ref{4.5.2}),(\ref{4.5.3}),(\ref{4.5.4})
imply that the representation on ${\cal H}_{+phys}^{*}$ is unitarily equivalent to an 
$L^2 (R, dx )$ representation. Specifically,
let $U$ be a linear map from ${\cal H}_{+phys}^{*}$ to $L^2 (R, dx )$ generated by its action 
on the positive frequency eigenstates as follows.
\begin{eqnarray}
U (\psi^{+}_{\frac{|k|}{\beta_0}} (l, \phi )
&=& \beta_0^{-\frac{1}{2}}e^{i \frac{|k|}{\beta_0}\phi}e^{-ikx}, \;\; k>0 ,
\label{upsiplus}\\
U (\psi^{-}_{\frac{|k|}{\beta_0}} (l, \phi )
&=& \beta_0^{-\frac{1}{2}}e^{i \frac{|k|}{\beta_0}\phi}e^{-ikx}, \;\; k<0.
\label{upsiminus}
\end{eqnarray}
This implies that 
\begin{eqnarray}
U (\psi (l, \phi ) =:\Psi (x, \phi ) 
&=& \int_{-\infty}^{\infty} dk f(k) e^{i \frac{|k|}{\beta_0}\phi}e^{-ikx} 
\label{upsi}\\
f(k) &=& \beta_0^{-\frac{1}{2}}f_+ (\frac{k}{\beta_0}) , \;\;k>0 \nonumber\\
&=& \beta_0^{-\frac{1}{2}}f_- (\frac{-k}{\beta_0}) , \;\;k<0 .
\label{uf}
\end{eqnarray}
The inner product on $L^2(R, dx )$ is 
\begin{equation}
(\Psi_1(x, \phi ), \Psi_2 (x, \phi )) 
=\int_{-\infty}^{\infty} dx \Psi_1^*(x, \phi ) \Psi_2 (x, \phi ).
\label{l2ip}
\end{equation}
It is easy to check that $U$ is a unitary map and that 
${\hat p}_{\phi}, {\hat x}_{\phi_0}$ act on states in $L^2(R, dx )$ via $U$ as
\begin{eqnarray}
{\hat p}_{\phi} \Psi (x, \phi ) &=& \frac{\hbar}{i} \frac{d}{d\phi} \Psi (x, \phi )
\label{pphil2}\\
{\hat x}_{\phi_0}\Psi (x, \phi )
&=& \int_{-\infty}^{\infty}dk(\frac{1}{i}\frac{d}{dk}+ \frac{k}{\beta_0|k|}\phi_0) f(k) 
e^{i \frac{|k|}{\beta_0}\phi}e^{-ikx}, 
\label{xphi0l2}
\end{eqnarray}
where in (\ref{xphi0l2}) $f(k)$ satisfies the appropriate conditions implied by
equation (\ref{hatxphi0domain}).

\section{The physical state space in the absence of Thiemann- like inverse triad definitions.}
Set $B(\mu )= |\mu |^{-3/4}$. Equation (\ref{defl}) implies that 
$B^{1/2} (\mu (l) ) = \frac{4}{3}|l|^{-1}$. Set $L>0$ in equation (\ref{defx}). Then we have that
\begin{eqnarray}
x(l) = \frac{4}{3} \int_L^l |{\bar l}|^{-1}&=&  \frac{4}{3} \ln\frac{l}{L} \nonumber\\
&=& \infty \; {\rm for}\; l\leq 0
\end{eqnarray}
The square integrability of $\psi (\phi )$ in (\ref{psil=l0}) implies that $\psi (\phi \pm (x(l_0)-x(l)))=0$
for $l\leq 0$. Thus for $l_0>0 $ the orbit of $\langle l_0 |$ under group averaging is effectively
$\{ \langle l |, l>0 \}$. To access $l<0$ we must choose $l_0<0$.
This implies that ${\cal H}_{phys}$ splits up into 2 orthogonal sectors, ${\cal H}_{>phys},{\cal H}_{<phys}$,
one obtained for $l_0>0$ and the other from $l_)<0$ (orthogonality follows from the 
inner product defined through group averaging).

Accordingly set $l_0=L=1$ to obtain ${\cal H}^*_{>phys}$ and $l_0=L=-1$ to 
obtain ${\cal H}^*_{<phys}$. It is then straightforward to see that the considerations of sections 
4.2- 4.4 apply to each sector ${\cal H}^*_{>phys}$,${\cal H}^*_{<phys}$,
 individually by virtue of the invertibility of $x (l)$ in each sector.
We shall use obvious notation with subscripts $>,<$ referring to the appropriate sector. Thus, we have that
\begin{eqnarray}
x(l):= x_{>}(l) &=& \ln |l| , l>0 \\
x(l):= x_{<}(l) &=& -\ln |l| , l<0 
\end{eqnarray}
so that $x_>, x_<$ are invertible functions of $l$ for $l>0,l<0$.

The positive frequency eigenfunctions now acquire a 4- fold (rather than 2- fold) degeneracy. These eigenfunctions,
 in the $l,\phi $ representation, are:
\begin{eqnarray}
\psi_{>\omega}^{\pm}(l, \phi ) &=& e^{i\omega \phi} e^{\mp \beta_0 \omega x_>(l)},\; l>0, \nonumber\\
&=& 0,\; l\leq 0 \\
\psi_{<\omega}^{\pm}(l, \phi ) = e^{i\omega \phi} e^{\mp \beta_0 \omega x_>(l)},\; l<0 ,\nonumber \\
&=& 0, \; l\geq 0 .
\end{eqnarray}
The representation on ${\cal H}_{+phys}^*$ is now unitarily equivalent to one on
$L^2(R, dx_>)\oplus L^2(R, dx_<) $. Replacing $x_>, x_<$ by $x_>(l), x_< (l)$ it is straightforward to 
see that 
\begin{equation}
L^2(R, dx_>)\oplus L^2(R, dx_<) = L^2 (R, |l|^{-1}dl ).
\end{equation}

The explicit unitary mapping is as follows.
A  positive frequency state $\psi (l, \phi )\in {\cal H}_{+phys}^*$ is characterised by its mode functions
$f_{>\pm} (\omega ), f_{<\pm}(\omega ), \omega >0$. Define the linear map $U$, 
$U:{\cal H}_{+phys}^*\rightarrow L^2 (R, |l|^{-1}dl )$ as follows.
\begin{eqnarray}
U (\psi (l, \phi ) =:\Psi (x, \phi ) 
&=& \int_{-\infty}^{\infty} dk f_>(k) e^{i \frac{|k|}{\beta_0}\phi}e^{-ik\ln|l|}, \; l>0 ,\label{76}\\
&=& \int_{-\infty}^{\infty} dk f_<(k) e^{i \frac{|k|}{\beta_0}\phi}e^{ik\ln|l|}, \; l<0 ,\label{77}
\end{eqnarray}
where the mode coefficients $f_>(k), f_< (k)$ are related to $f_{>\pm} (\omega ),f_{<\pm}(\omega )$ through
the analogs of equation (\ref{uf}) i.e. $f(k), f_{\pm} (\omega )$ in that equation are replaced by 
$f_>(k),f_{>\pm} (\omega )$ and $f_<(k),f_{<\pm} (\omega )$.

The inner product between $\Psi_1(l, \phi ), \Psi_2 (l, \phi )\in L^2(R,|l|^{-1}dl )$ is 
\begin{equation}
(\Psi_1(x, \phi ), \Psi_2 (x, \phi )) 
=\int_{-\infty}^{\infty} \frac{dl}{|l|} \Psi_1^*(l, \phi ) \Psi_2 (l, \phi ).
\label{wdwl2ip}
\end{equation}
It is straightforward to check that $U$ is a unitary map and 
that 
${\hat p}_{\phi}, {\hat x}_{\phi_0}$ act on states in $L^2(R,|l|^{-1} dl )$ via $U$ as
\begin{eqnarray}
{\hat p}_{\phi} \Psi (x, \phi ) &=& \frac{\hbar}{i} \frac{d}{d\phi} \Psi (x, \phi ),
\label{wdwpphil2}\\
{\hat x}_{\phi_0}\Psi (x, \phi )
&=& \int_{-\infty}^{\infty}dk(\frac{1}{i}\frac{d}{dk}+ \frac{k}{\beta_0|k|}\phi_0) f_>(k) 
e^{i \frac{|k|}{\beta_0}\phi}e^{-ik\ln l},\;l>0 ,
\label{1wdwxphi0l2}\\
&=& \int_{-\infty}^{\infty}dk(\frac{1}{i}\frac{d}{dk}+ \frac{k}{\beta_0|k|}\phi_0) f_<(k) 
e^{i \frac{|k|}{\beta_0}\phi}e^{ik\ln |l|},\;l<0 ,
\label{2wdwxphi0l2}
\end{eqnarray}
where in (\ref{1wdwxphi0l2}),(\ref{2wdwxphi0l2}), $f_>(k), f_< (k)$ satisfy the appropriate conditions
implied by replacing $f_{\pm }(\omega )$ by $ f_{>\pm }(\omega ),f_{<\pm }(\omega )$ 
on equation(\ref{hatxphi0domain}). Thus ${\hat x}_{\phi_0}$ is defined on a dense domain wherein
$f_>(k), f_<(k)$ and all their derivatives vanish at $k=0$ and $|k|\rightarrow \infty$.

The resultant representation on 
$L^2(R, |l|^{-1}dl )$ is exactly the Wheeler DeWitt representation of \cite{aps2}. As in 
that work, the operator ${\hat \Pi}$ defined by ${\hat \Pi}|l\rangle = |-l\rangle$ is a 
large gauge transformation . This implies that physical positive frequency states lie in the 
symmetric sector of ${\cal H}_{+phys}^*$ whose image under $U$ is the symmetric sector
of $L^2(R, |l|^{-1}dl )$ characterised by 
\begin{equation}
F(k):=f_>(k) = f_<(-k) .
\label{symmwdw}
\end{equation}
This completes our analysis of the case wherein we drop (iii) of section 1.

Before we proceed to the next section  wherein (iii) is retained
we would like to stress an important technical issue overlooked by APS in their analysis of WDW states.
It is very important that the domain of the operator ${\hat x}_{\phi_0}$ be defined carefully as
we have done above. The wave functions in this domain are in Schwartz space i.e  they fall off faster 
than any power of $\ln |l|$. 

APS require that wave functions have Schwartz space behaviour in 
$\ln |l|$ (actually they require such behaviour in $l$ which is difficult to implement)
 and they equate this requirement with Schwartz space behaviour of 
$F(k)$ in equation (\ref{symmwdw}). However this 
is incorrect due to the presence of the $e^{i\frac{|k|}{\beta_0}\phi}$
term in the mode expansions. This term is not differentiable with respect to $k$ at $k=0$ and
if $F(k)$ does not satisfy our requirements at $k=0$, the resulting wave function has a power law
fall off at infinity \cite{meinprep}.  Further, APS choose  ${\hat \mu}_{\phi_0}$ as their Dirac 
observable. Since this is exponentially related to ${\hat x}_{\phi_0}$, this operator does not have
well defined action on wave functions for which $F(k)$ is merely Schwartz. 
This has important implications for their semiclassical analysis. APS choose $F(k)$ to be 
a Gaussian - hence it is not in our domain. In their evaluation of  expectation values and fluctuations
of ${\hat {|\mu |}}_{\phi_0}$ they neglect certain terms which, due to bad infrared behaviour in $\ln |l|$
contribute divergently \cite{meinprep}. As we shall show in a subsequent paper \cite{meinprep}, the expectation value and 
fluctuations of ${\hat x}_{\phi_0}$ do exist in this state and behave reasonably.
However, these technicalities are not expected to find their way into their LQC `quantum bounce' results. The reason
is that, amongst other checks, APS have evaluated the wave function in $l,\phi$ space by evaluating the 
relevant integral (see equations (\ref{76}),(\ref{77}),(\ref{symmwdw}))
in $k$- space numerically and in their numerical evaluation
they have used a function which is Gaussian near its peak but which is of compact support in $k$ and whose support 
is outside $k=0$.

\section{The physical state space in the presence of Thiemann- like inverse triad definitions.}
Let $B(\mu )$ be obtained from a Thiemann like prescription \cite{ttinverse,martinreview} and set $l_0=L=0$.
Since $\frac{dx}{dl}= B^{1/2}( \mu (l) )$ and $B$ obtained in \cite{martinreview,aps1} or \cite{aps2} is
positive and non-vanishing except at $l=0=x(l=0)$, it follows that $x$ is an invertible function of $l$.
Hence all the considerations of section 4 apply. In particular, the positive frequency eigen functions are
doubly degenerate unlike in the WDW case. Symmetric states in ${\cal H}^*_{+phys}$ or, equivalently, in 
$L^2(R, dx )$ are defined by $f_+ (\frac{|k|}{\beta_0})=f_- (\frac{|k|}{\beta_0})$ or, equivalently,
$f(k)= f(-k)$ (see equation (\ref{uf})).

From equations (\ref{53}) and (\ref{54}) it follows that 
\begin{equation}
\frac{d{\hat x}_{\phi_0}}{d\phi_0}= \pm \beta_0^{-1}.
\label{eomxhat}
\end{equation}
Thus implies that the corresponding classical evolution equations with respect to the scalar field `time'
are
\begin{equation}
\frac{dx(\phi )}{d\phi}= \pm\beta_0^{-1}
\label{eomx}
\end{equation}
which have the solutions
\begin{equation}
x-x_* = \pm \beta_0^{-1}(\phi -\phi_*)
\label{xsoln}
\end{equation}
The expanding branch follows  $x-x_* =\beta_0^{-1}(\phi -\phi_*)$ so that evolving backwards 
from some large $x_*$, the value $x=0=\mu$ is reached in finite scalar field- time. Thus the epoch when the 
size of the universe
vanishes is reached in finite time. However, \\
\noindent (a) the scalar field energy density, $\rho =(\frac{8 \pi \gamma l_P^2}{6})^3(p_{\phi}B(\mu ))^2$
is bounded throughout and vanishes at $\mu =0$.\\
\noindent (b) there is regular evolution beyond the point at which the size of the universe vanishes.

In this sense the singularity is resolved. In the next section we discuss the freedom in defining $B$ and 
argue that a choice of $B(\mu )$ exists for which the phenomenon of singularity resolution is independent of the 
choice of the fiducial cell ${\cal V}$ which underlies the quantization (see section 2).

\section{Is singularity resolution without curvature corrections  physically well defined?} 

If we use the available choices of $B (\mu )$ in the literature \cite{martinreview,aps1,aps2}, the following 
unphysical situation (pointed out by APS in \cite{aps2})  is encountered. Typically $B(\mu )$ departs signficantly
from the WDW choice $|\mu|^{-3/2}$ when $\mu$ is close to some fixed $\mu_0$ \cite{martinreview,aps1}
or when $v= {\rm sgn}(\mu )\mu^{3/2}$ is close to 1 \cite{aps2}. This is the stage at which quantum effects
become important. However, $(\frac{8 \pi \gamma l_P^2}{6})^{3/2}|v|$ is the volume of the 
fiducial cell ${\cal V}$ with fiducial volume $V_0$. Since the choice of ${\cal V}$
is arbitrary, physical phenomena, such as the stage at which quantum effects become important,
 should not depend on this choice and therefore singularity resolution in the 
context of such a choice of $B$ cannot be taken seriously.

The choice of $B$ in the APS work \cite{aps1,aps2} is motivated by requiring that the physical length scale, associated
with the loop which labels the holonomy which regulates the definition of ${\hat{|p|}}^{-3/2}$,  be 
of order of the Planck length. This motivation, the consequent choices of holonomy operator in \cite{aps1,aps2} 
and the arguments mentioned above against the
physical significance of singularity resolution due to inverse triad definitions, form one consistent viewpoint.
Below we propose a different viewpoint, also self consistent, which motivates an alternate choice of $B$.

The LQC quantization requires a choice of fiducial metric and fiducial cell. These are auxilliary structures
which facilitate the explicit imposition of homogeneity and isotropy and which allow the definition of 
spatial integrals, vital to the Hamiltonian framework, which would diverge if evaluated over non- compact 
spatial manifold. Their auxilliary nature requires, as mentioned above, that physical results do not depend
on the particular choice. Let us first
fix the fiducial metric (we shall relax this later). 
Then the  only choice in the quantization is that of the fiducial 
cell. Denote two such choices by ${\cal V}_1,{\cal V}_2$ with fiducial volumes $V_{0,1},V_{0,2}$. Fix  the 
state $|v_1\rangle_1$ ($v$ is related to $\mu$ as described above; 
$(\frac{8 \pi \gamma l_P^2}{6})^{3/2}v$ is the eigen value of the volume operator,
${\hat{|p|}^{3/2}}$)  in the first case. Then the physical volume of the cell ${\cal V}_1$  is 
$(\frac{8 \pi \gamma l_P^2}{6})^{3/2}v_1$. Similarly the physical volume of the cell
${\cal V}_2$  in the state $ |v_2\rangle_2$
is $(\frac{8 \pi \gamma l_P^2}{6})^{3/2}v_2$. Hence the physical volume of ${\cal V}_1$ in the 
state $ |v_2\rangle_2$ is $\frac{v_2}{V_{0,2}}V_{0,1}$. Hence the states
$|v_1\rangle_1$, $ |v_1 \frac{ V_{0,2} }{ V_{0,1} }\rangle_2$
describe the same physical spatial geometry. Hence the inverse volume of ${\cal V}_1$ in these
states should also be identical. Denoting the inverse volume functions (modulo the factor 
$(\frac{8 \pi \gamma l_P^2}{6})^{3/2}$) in the 2 cases by $B_1(v), B_2(v)$, the above discussion implies that 
\begin{equation}
B_1( v_1 ) = \frac{ V_{0,2} }{ V_{0,1} } B_2 (\frac{ V_{0,2} }{ V_{0,1} }v_1).
\label{7.1}
\end{equation}
This is not satisfied by the choices of $B$ in \cite{martinreview,aps1,aps2}. It is, however, 
straightforward to check that if we replace the APS \cite{aps2} holonomy operators
$\widehat{e^{\frac{i{\bar \mu}c}{2}}}$ (see \cite{aps2} for a definition of $\bar \mu$ and of this operator) by 
$\widehat{e^{\frac{i\lambda_1{\bar \mu}c}{2}}}$ in case 1 and 
$\widehat{e^{\frac{i\lambda_2{\bar \mu}c}{2}}}$ in case 2 
then we have that for $i=1,2$
\begin{equation}
B_i( v) = (\frac{3}{2})^3K|v|\{ \frac{|v+\lambda_i|^{1/3}- |v-\lambda_i|^{1/3}}{\lambda_i} \}^3,
\label{biv}
\end{equation}
whre $K$ is a numerical factor defined in \cite{aps2}.
If we set 
\begin{equation}
\frac{\lambda_1}{\lambda_2}= \frac{ V_{0,1} }{ V_{0,2} }
\label{lambdav}
\end{equation}
then it follows straightforwardly that equation (\ref{7.1}) holds. The choices
(\ref{biv}), (\ref{lambdav}) not only yield the correct scaling (\ref{7.1}) 
but mesh well with an alternate viewpoint on the significance of the holonomy operator, pointed out 
in \cite{ttsingavoid}. The authors of that work note that the inverse triad is obtained classically by a 
derivative of the volume function $V$.  In the LQG regularization this is replaced by a ``discrete derivative''
\cite{ttsingavoid} through the structure ${\hat V}- {\hat h}^{-1}{\hat V}{\hat h}$ where $\hat h$ is the holonomy
operator. Indeed, in LQC, the APS holonomy operator \cite{aps2} provides {\em exactly} such a realization of
the regularization process by virtue of its being a displacement operator for $|v \rangle$. Viewed in this light,
consider, once again the APS choice \cite{aps2} for ${\hat h}$. For the quantization based on the fiducial cell
${\cal V}_1$ this operator increments the physical volume of ${\cal V}_1$ by the fixed amount 
$(\frac{8 \pi \gamma l_P^2}{6})^{3/2}$. For the quantization based on ${\cal V}_2$, the operator 
increments the physical volume of ${\cal V}_2$ by this same amount and consequently increments the physical volume of
the cell ${\cal V}_1$ by the amount 
$(\frac{8 \pi \gamma l_P^2}{6})^{3/2}\frac{ V_{0,1} }{ V_{0,2} }$. If we demand that the volume displacement 
be independent of the choice fiducial cell ${\cal V}$ we again arrive at the modified operators
$\widehat{e^{\frac{i\lambda_1{\bar \mu}c}{2}}}$,  
$\widehat{e^{\frac{i\lambda_2{\bar \mu}c}{2}}}$ with $\lambda_1, \lambda_2$ satisfying equation (\ref{lambdav}).

This implies that that there is some region $R$ of fiducial size $V_0(R)$ for which the physical volume increment
is exactly $(\frac{8 \pi \gamma l_P^2}{6})^{3/2}$ {\em independent} of the choice of ${\cal V}$ so that 
$\lambda_i = \frac{V_{0,i}}{V_0(R)}, i=1,2$. At first sight this seems unphysical.
Note however that the spatial diffeomorphism gauge freedom is fixed so that the fiducial coordinates are of
physical, gauge invariant relevance even though they are not metrical distances.
The above discussion is predicated on a fixed choice of fiducial metric consistent with homogeneity and
isotropy. However, as we describe below, our considerations are independent of this choice.

Our picture is as follows. 
Underlying the effective description provided by LQC, is a physical state of the full 
theory. The full theory has gravity, scalar field matter as well as suitable matter degrees of freedom
which define physical spatial coordinates in a manner envisaged by, for example, Rovelli in \cite{rovelli}.
 Thus the  underlying LQG state has enough structure to
 enable the definition of a class of physical coordinate systems 
on the spatial manifold $\Sigma$ such that homogneity and 
isotropy are manifest. Isotropy and homogeneity imply that
these physical coordinate systems are related to
 each other by constant rescaling. These are just the set of all comoving
 coordinate systems.
 
 Homogeneity  itself is expected to arise as a good effective property
 after averaging out microphysics.  The scale at which homogeneous modes
 are a sufficiently good description is time dependent- for example, a very
 large scale today is much smaller at an earlier epoch if the two epochs are part of an
 expanding phase. A time independent notion of the minimal domain
 wherein a homogeneous description is adequate can be stated in terms of
 comoving coordinates provided by the underlying LQG state as follows.
  Around any point $p \in \Sigma$, the LQG state defines a minimal
 region $R_p$ wherein homogeneity is a good effective description.
 Homogeneity requires that $R_p$ takes the form of  a $p$ independent
 comoving cell around $p$. This is exactly the cell $R$ alluded to above.
 
 The choice of comoving coordinates naturally defines a flat fiducial metric.
 Different choices endow the cell $R$ with different fiducial sizes.
 Note that while the fiducial volume of $R$ depends on the choice
 of comoving coordinates, $R$ itself is independent of this choice.
 
As indicated above,
 LQC requires a choice of  cell (in general different from $R$)
 as well as a fiducial metric. As we shall see explicitly below, 
introducing the $\lambda$ dependent holonomies (see the discussion before
(\ref{biv})) ensures that 
 inverse volume effects become important when the {\em physical}  volume of  the cell $R$ approaches 
$(\frac{8 \pi \gamma l_P^2}{6})^{3/2}$.
 This statement is independent of the LQC choice of cell and fiducial metric.
We have already shown that the statement holds for a fixed fiducial metric. We now show that 
it also holds when we vary the choice of (comoving coordinates and associated) fiducial metric.

We define the following notation.
For the LQC quantization choice of cell ${\cal V}_k$ and comoving coordinate system $\{x\}_{\alpha}$,
denote the  fiducial volume of the cell by $V_{\alpha , k}$
and the volume eigenstates by $|v\rangle_{\alpha , k}$ so that $(\frac{8 \pi \gamma l_P^2}{6})^{3/2}v$ is the 
physical volume of ${\cal V}_k$. 
Arguments identical to those for the case of fixed choice of fiducial metric, indicate that, in obvious notation,
 we must identify the state
$|v_{\alpha, k}\rangle_{\alpha , k}$ with the state $|v_{\beta , l}\rangle_{\beta , l}$ iff the eigenvalues 
$v_{\alpha , k},v_{\beta , l}$ satisfy the relation
\begin{equation}
\frac{v_{\beta , l}}{v_{\alpha ,k}}= \frac{V_{\beta , l}}{V_{\beta ,k}}=\frac{V_{\alpha , l}}{V_{\alpha ,k}},
\label{7.2}
\end{equation}
where we have used the fact that the $\alpha$ and $\beta$ fiducial metrics are related by constant rescaling.
The above equation shows that the identification of states is independent of the 
fiducial metric label ${\alpha}$ so that we may drop the $\alpha, \beta$ labels from the eigen values
$v_{\alpha ,k},,v_{\beta , l}$ and denote the identified pair of states
by $|v_{k}\rangle_{\alpha , k}, |v_{l}\rangle_{\beta , l}$ with 
\begin{equation}
\frac{v_{l}}{v_{k}}= \frac{V_{\gamma , l}}{V_{\gamma ,k}} \forall \gamma,
\label{7.3}
\end{equation}
where,  once again, we have used the fact that all fiducial metrics are 
related by constant rescalings.

Next, denote the holonomy operators appropriate to the two sets of choices by.
$\widehat{e^{\frac{i\lambda_{\alpha , k}{\bar \mu}c}{2}}}$,
$\widehat{e^{\frac{i\lambda_{\beta , l}{\bar \mu}c}{2}}}$. 
The action of the operators
$\widehat{e^{\frac{i\lambda_{\alpha , k}{\bar \mu}c}{2}}}$,
$\widehat{e^{\frac{i\lambda_{\beta , l}{\bar \mu}c}{2}}}$ on the states 
$|v_{k}\rangle_{\alpha , k}, |v_{l}\rangle_{\beta , l}$, yield the states
$|v_{k}+\lambda_{\alpha , k}\rangle_{\alpha , k}, |v_{l}+\lambda_{\beta , l}\rangle_{\beta , l}$
In analogy to our arguments for the 
fixed fiducial metric case, we require that these operators 
increment the {\em physical} volume of $R$ by the fixed amount 
$(\frac{8 \pi \gamma l_P^2}{6})^{3/2}$. It is easy to check that this implies that 
\begin{equation}
\lambda_{\alpha , k} = \frac{v_k}{v_{R}},\;\;
\lambda_{\beta , l} = \frac{v_l}{v_{R}}
\label{7.4}
\end{equation}
where $(\frac{8 \pi \gamma l_P^2}{6})^{3/2}v_R$ denotes the physical volume of the 
region $R$ in the state 
$|v_{k}+\lambda_{\alpha , k}\rangle_{\alpha , k}$. This volume is identical to that in the 
state $|v_{l}+\lambda_{\beta , l}\rangle_{\beta , l}$ by virtue of equation (\ref{7.3}).
Thus the $\lambda$ factors are independent of the fiducial metric. This implies that the inverse
volume functions (in obvious notation) $B_{\alpha , k}, B_{\beta , l}$ are independent of the 
fiducial metric labels $\alpha ,  \beta$.
This completes the argument.

\section{Discussion}

The simplicity of the Hamiltonian constraint for spatially flat, homogeneous and isotropic gravity  
coupled to a massless scalar field allows us to define its exponentiated square roots as unitary operators
on the standard LQC kinematic Hilbert space \cite{martin} using a technique introduced in Reference \cite{aps2}. 
This allows the application of group averaging methods
coupled with some key ideas (see section 4) to construct the physical Hilbert space of the model without any
of the higher order curvature corrections which occur in all previous LQC treatments of the model. This allows 
us to evaluate the role of various exotic features of LQC in the phenomenon of singularity resolution in the model.

As expected, from prior work \cite{apsrobust,martineffective}, our work confirms that the quantum bounce of the 
APS quantizations \cite{aps1,aps2} is clearly due to the higher order curvature corrections (see (ii) of the Introduction)
to general relativistic dynamics which are present in their work. In the absence of such corrections we have demonstrated
in sections 5 and 6 that the bounce does not occur. Remarkably, without such corrections, the kinematic discreteness
of the LQC representation does not leave any imprint on our physical Hilbert space representations. Thus the 
physical Hilbert space representations of sections 5 and 6 do not split into an uncountable number of 
seperable superselection sectors each characterised by a 1 dimensional regular lattice
coordinatised by discrete values of the $\mu$ \cite{aps1} or $v$ \cite{aps2} variables. Instead, the physical Hilbert
space obtained in this work, by group averaging, is always seperable and the variable $x$ of section 6 ranges over
the entire real line.
\footnote{Appendix C of Reference \cite{aps1} details an $L^2(R)$ physical Hilbert space representation; however the 
representation still has traces of the kinematic discreteness characteristic of LQC by virtue of the role of
a {\em difference} operator in the definition of quantum dynamics and Dirac observables.}

If in addition to a curvature correction free LQC quantization of the model, one uses the straightforward 
spectral analysis based inverse scale factor operator to define the matter density, the physical
Hilbert space representation turns out to be  the WheelerDeWitt one. If one uses Thiemann- like 
\cite{ttinverse,martin,aps1,aps2}
definitions for the inverse scale factor operator the physical state representation is inequivalent to the 
WDW one and exhibits singularity resolution through well defined, regular (backward) evolution through the 
classically singular region. Such singularity resolution is physically unambiguous only if we slightly alter the 
definition of the APS ``improved'' inverse volume operator of Reference \cite{aps2} along the lines indicated
in section 7. Remarkably, 
the scaling  requirements of section 7 {\em cannot} be satisfied for the triad independent holonomy operator
of Reference \cite{aps1}. Also remarkable, is that the scaling requirements {\em are} satisfied for all 
choices of Bojowald's ambiguity parameter $l$ (at least for the $j=1/2$ case) \cite{martin} provided that 
his expressions are suitably generalised to the context of the triad dependent operators introduced
in section 7.

As mentioned above, our considerations are tied to the simplicity of the Hamiltonian constraint of the model. 
While it is conceivable that our considerations may generalise to the spatially closed isotropic  model (with a massless
scalar field as in this work), we do not see how to apply our ideas to more complicated settings
in which anisotropies play a role such as in the homogeneous diagonal models of Reference \cite{martindiagonal}.
Thus, in the context of more general settings our results here can, at best, be viewed as suggestive of the nature of 
quantum inverse triad effects near singularities. 

In our work, singularity resolution only depends on quantum effects which become important when the 
physical size of the region $R$ (see section 7) becomes of the order of the Planck volume (assuming that the 
Immirzi parameter is of order unity). The mechanism of singularity resolution is thus independent of the matter density. Thus, the  universe at large
size could have an arbitrarily large matter density (by choosing $p_{\phi}$ to be arbitrarily large) and
still behave classically. This is in contrast to the APS mechanism of singularity resolution in Reference \cite{aps2}
wherein the density is always bounded by the critical density at the bounce if the state at large volume
is to be semiclassical. It would be of interest to compare our viewpoint (see section 7) on the physical validity of 
inverse volume operator
driven 
singularity resolution with  
the ``lattice refinement'' picture of Bojowald \cite{martinlattice}, particularly the relation between the region $R$
here and the lattice parameter $l_0$ in Reference \cite{martinlattice}.

\vspace{2mm}

\noindent
{\bf Acknowledgements}: I thank Alok Laddha for very useful discussions. I am very grateful to
 Abhay Ashtekar and Martin Bojowald for their comments on a preliminary version of this work. I thank
Ghanashyam Date for inviting me to the LQC August 2008 meeting in IMSc, Chennai, where this work was initiated.

\end{document}